\documentclass[conference]{IEEEtran}

\usepackage{epsfig}
\usepackage{epstopdf}
\usepackage{amssymb}
\usepackage{amsmath}
\usepackage{amsfonts}
\usepackage{subfigure}

\DeclareFontFamily{OT1}{pzc}{}
\DeclareFontShape{OT1}{pzc}{m}{it}{<-> s * [1.10] pzcmi7t}{}
\DeclareMathAlphabet{\mathpzc}{OT1}{pzc}{m}{it}

\begin{document}
\title{Social Ranking Techniques for the Web}

\author{\IEEEauthorblockN{Tommy H. Nguyen}
	\IEEEauthorblockA{Rensselaer Polytechnic Institute\\
	110 8th St, Troy, NY 12180 \\
	Tel/Fax: (518) 276-\{2094,2529\} \\
	Email Address: nguyet11@rpi.edu}
\and
	\IEEEauthorblockN{Boleslaw K. Szymanski}
	\IEEEauthorblockA{Rensselaer Polytechnic Institute\\
	110 8th St, Troy, NY 12180 \\
	Tel/Fax: (518) 276-\{2716,2529\} \\
	Email Address: szymab@rpi.edu \\ }}

\maketitle
\begin{abstract}
The proliferation of social media has the potential for changing the structure and organization of the web. In the past, scientists have looked at the web as a large connected component to understand how the topology of hyperlinks correlates with the quality of information contained in the page and they proposed techniques to rank information contained in web pages. We argue that information from web pages and network data on social relationships can be combined to create a personalized and socially connected web. In this paper, we look at the web as a composition of two networks, one consisting of information in web pages and the other of personal data shared on social media web sites. Together, they allow us to analyze how social media tunnels the flow of information from person to person and how to use the structure of the social network to rank, deliver, and organize information specifically for each individual user. We validate our social ranking concepts through a ranking experiment conducted on web pages that users shared on Google Buzz and Twitter. 
\end{abstract}

\section{Introduction}
According to \cite{Kleinberg:2001}, a conceptualization of the web is revealed by looking at patterns in the topology of hyperlinks containing web pages to separate prominent websites that serve as authorities for trusted information from malicious pages created by spammers. This conceptualization of the web eliminates the complexity of textual analysis and creates a pot-pourri of information that gets incorporated into search engines for the purpose of finding information on computing devices. 

Advances in social networks have provided a new dimension to studying problems in information retrieval from a network point of view. Incorporating the social network structure into algorithms used for ranking, organizing, and delivering information in information retrieval systems such as search engines have promising improvements and new practical applications. For example, ``movies that my friends like" has been introduced by Facebook as graph search. 

The advances of the web have created some applications where humans can identify and label relationships for the purpose of interacting with information. Beside the typical information that users share in online social networks such as photos, messages, geographic locations, etc., URLs that users share with their friends and followers are used in this paper to infer how humans would rank the importance of the content embedded on the page because URLs shared by users focus on selected topics that they want their followers to know. Therefore, publicly shared messages embedded with URLs provide us a clue into how a user would rank the importance of a page, which defines the ranking of the page by the view of the user, and allow us to re-rank, re-organize, and re-deliver query results based on who is connected to whom.

We propose techniques for answering the following questions. First, how can we incorporate social relevance into the process of ranking pages while preserve authoritative sources determined by algorithms based on indegree analysis such as PageRank and HITS? Second, how can we rank pages based on URLs that users share in online social
media such as Google Buzz and Twitter by incorporating the social network structure of those users to personalize the ranking of pages tailored to each individual user?

\begin{figure}
	\center
	\includegraphics[scale=0.60]{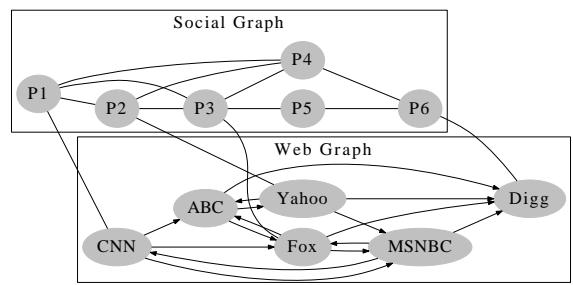}
	\caption{Previous work on the ranking of pages conceptualized the web as a network consisting of pages representing nodes, and links representing directed edges illustrated in the bottom rectangular box. Advances in social networks enabled a different perspective of the ranking of pages studied in this paper. For simplicity, the social network of users illustrated in the top rectangular box consisting of nodes $P_1, P_2, ..., P_6$ where an undirected edge between $P_1$ and $P_2$ represents a social relationship between the two nodes and an undirected edge from $P_{1}$ to CNN represents $P_{1}$ broadcasting a CNN URL to its ties $P_{2}, P_{3}, P_{4}$. Note that the edge from $P1$ to CNN is not a part of the social network, but connection between the web and social network.}
\end{figure}

The rest of the paper is organized as follows. In section \ref{algorithms}, we provide techniques for ranking pages by applying PageRank, HITS, and maximum flow to social ties and URL-embedded messages shared on social media. In section \ref{data-collection}, we overview the procedure for collecting data on two social media (Google Buzz and Twitter) for the validation of our proposed framework by ranking URLs shared in them. In section \ref{search_experiment}, we analyze the social relevance of URLs and conduct a ranking experiment to observe the ranking positions of URLs computed by PageRank, HITS, and maximum flow. After presenting the literature review of ranking and other related work in section \ref{related_work}, we conclude in section \ref{conclusion} by summarizing the results.

\section{Social Ranking Techniques}
\label{algorithms}
Let $G_U=(V,E)$ be a directed multi-labeled graph where $V$ is the set of nodes, $E$ is the set of edges where $e=(v_i, v_j)$ represents a directed edge from node $v_i$ to node $v_j$, and $U$ is the set of URLs with subsets of which nodes in $V$ are labeled. For URL $u \in U$, let $S(u)$ denotes the set of all spreaders of the URL $u$; in other words all nodes in $V$ who has posted $u$. 

\subsection{PageRank on Social Network (PRSN)}
We extend the PageRank algorithm to rank URLs on a social network (PRSN) as follows. Given a multi-labeled graph $G_U=(V,E)$, let $F=(f_{ij})$ be a $n \times n$ weighted adjacency matrix where $n$ is the number of nodes (i.e, $n=|V|$), $f_{ij}=0$ if there is no directed edge from $v_i$ to $v_j$, and $f_{ij} = 1 / deg(i)$ otherwise. Let $R$ be a vector consisting of $n$ elements where the $i^{th}$ element of $R$ denoted as $r_i$ corresponds to the PageRank score of the $i^{th}$ node. Let $k$ be the maximum number of iterations that the PageRank algorithm runs. At the first iteration, every node sends its score divided by the number of links pointing from this node to other nodes through each outgoing link. Then each node updates its score to the sum of scores that it has received; that is, 

\begin{eqnarray}
	\label{eq:pgrank1}
	r_i &=& f_{1i}r_1 + f_{2i}r_2 + ... + f_{ni}r_{n}
\end{eqnarray}

If there is an edge from node $j$ to node $i$, then $f_{ji} > 0$ and node $j$ will send $f_{ji}$ fraction $\frac{1}{deg(j)}$ of its score $r_j$ to node $i$. Equation \ref{eq:pgrank1} can be compactly written as $R^{<1>}=F^{T}R^{<0>}$ where $F^{T}$ is the transpose of the matrix $F$, the superscript $^{<1>}$ denotes the scores of all nodes after the first iteration, and $R^{<0>}$ is the initial vector.  Let $R^{<k>}$ be the scores of nodes at the $k^{th}$ or last iteration defined as:

\begin{eqnarray}
	\label{eq:pgrank2}
	R^{<i>} = F^{T}R^{<i-1>} \;\;where\;\; 0 < i \leq k. 
\end{eqnarray}

If there are sinks in the graph $G$, that is nodes without outgoing edges, then for large enough $k$'s they will absorb all scores since the scores can enter but cannot leave the sinks. One way to fix this problem is to scale the strength of links by a constant factor of $0 < \sigma < 1$ and to compensate this scaling by adding an artificial flow between any two nodes with the weight $\frac{1-\sigma}{n}$. This solution is known as the scaled version of PageRank. The score of the $i^{th}$ node is then denoted as $r'_i$ and is defined as:

\begin{eqnarray}
	\label{eq:pgrank2a}
	r'_i &=& \sum_{j=1}^{n}(\sigma f_{ji} + \frac{1-\sigma}{n})r'_j. 
\end{eqnarray}

Equation \ref{eq:pgrank2} can be compactly written using the following matrix $\tilde{F}=\sigma F + \frac{1-\sigma}{n}$. By the Perron-Forbenius Theorem \cite{easley:2010}, the scaled PageRank scores converge to a stable solution:

\begin{eqnarray}
	\label{eq:pgrank3}
	R'^{i}=\tilde{F}^{T}R'^{i-1} \;\;where\;\; 0 < i \leq k.
\end{eqnarray}

Given a subset of URLs $U' \subset U$, the PageRank score of a URL $u \in U'$ on a social network (PRSN) is defined as:

\begin{equation}
	 PRSN(u) = \frac{\sum_{v_i \in S(u)}r_i'^k }{\sum_{u'\in U'} \sum_{v_i \in S(u')} r_i'^k}.
\end{equation}

\subsection{HITS on Social Network (HSN)}
The HITS algorithm used to rank URLs on a social network (HSN) is defined as follows \cite{easley:2010} \cite{Kleinberg-hits}. Given $G_U=(V,E)$, let $M=(m_{ij})$ be a $n \times n $ adjacency matrix where $n$ is the number of nodes, $m_{ij}=1$ if there is a directed edge from node $v_i$ to node $v_j$, and $m_{ij}=0$ otherwise. Let $k$ be the maximum number of iterations. Given a set of URLs $U' \subset U$, let $H$ and $A$ be vectors of scores for hubs and authorities, respectively. Authorities are the URLs (i.e., $u \in U'$) and hubs are nodes that share these URLs. The $i^{th}$ element of the vector $H$ represents the score of the $i^{th}$ hub, and the $j^{th}$ element of the vector $A$ represents the score of the $j^{th}$ authority. At the first iteration, the score $h_i$ of a hub gets set to the number of authorities to which it points, and the score $a_j$ of an authority gets set to the scores of hubs pointing to it. More formally, $h_i$ and $a_j$ are defined as: 

\begin{eqnarray}
	\label{hits:eq1}
	h_i^{<0>} &=& m_{i1}+m_{i2}+...+m_{in} \\
	a_j^{<0>} &=& m_{1j}h_1^{<0>} +m_{2j}h_2^{<0>} +...+ m_{nj}h_n^{<0>}
\end{eqnarray}

Let $H^{<l>}$ and $A^{<l>}$ be the scores of hubs and authorities at the iteration $l$, the HITS algorithm \cite{easley:2010} can be written as:

\begin{eqnarray}
	H^{<l>} &=& (MM^{T})^{l}H^{<0>} \;\; where \;\; 0 < l \leq k\\
	A^{<l>} &=& (M^{T}M)^{l-1}M^{T}H^{<0>} \;\; where \;\; 0 < l \leq k
\end{eqnarray}

Finally, the score of a URL in the authorities is the value $a_j^{<k>}$ normalized by the sum of scores in the vector $A$. 

\subsection{Social Ranking with Maximum Flow}
We defined the following maximum flow algorithm to rank URLs on a social network. Given a graph $G_U=(V,E)$ and a subset of URLs $U' \subset U$, let $p$ represent a node. We want to rank the URLs in $U'$ with respect to $p$ and $G$ by constructing a directed flow graph denoted as $G'_p=(V', E')$. 

The first part of the construction requires copying the social structure of $G$ to $G_{p}'$. For every node $v_i$ that $p$ follows, we add $v_i$ to $V'$ and the edge $e=(p, v_i)$ into $E'$. At the subsequent iteration, we repeat the same process for every node that has been added into $V'$ from the previous iteration; that is, if $v_i$ was added into $V'$ and there is an edge $e=(v_i, v_j)$, then we add $v_j$ to $V'$ if $v_j$ has not been added before. The edge $e=(v_i, v_j)$ will still be added into $E'$  if $v_j$ has been added before. This process of constructing the graph $G'_p$ continues until all possible nodes from $V$ that are reachable from $p$ have been added into $V'$. For practical reasons, it is wise to stop when the diameter of $G'_p$ is small; e.g., three to reflect the influence of nodes that are within network proximity. At the end of the process, an edge originating from node $v$ gets the weight equal to the inverse of the node degree in $G'_p$.

The second part of constructing $G'_p$ introduces some additional nodes and edges. For every URL $u' \in U'$, we add $u'$ into $V'$. For every spreader $s \in S(u')$ of the URL $u'$, we add an edge $e=(s,u')$ with a weight of 1 into $E'$ if $s \in V'$. We add a super sink denoted $t$ into $V'$ and add an edge $e=(u',t)$ with an edge weight of $1$ for every URL $u'$ in $U'$. 

The maximum flow of the graph $G_{p}'$ from source $p$ to super sink $t$ is a function $\mathpzc{F}$ that assigns a non-negative value to each edge so that it maximizes the total flow coming from the source $p$ to the super sink $t$ satisfying two conditions: first, it does not exceed the weight of an edge; i.e, $\mathpzc{F}(e) \leq c_{e}$  and second, it obeys the conservation of flow law except for the source $p$ and the super stink $t$; i.e, 

\begin{equation}
\mathpzc{F}_{out}(v) = \overbrace{\sum_{} c_{e}}^\text{Flow out to social ties} + \overbrace{\sum_{ } c^{'}_e}^\text{Flows out to pages} =  \mathpzc{F}_{in}(v)
\end{equation}

\begin{figure}
	\center
	\includegraphics[scale=0.70]{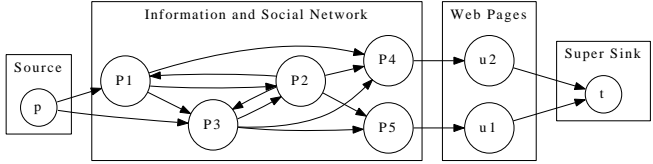}
	\caption{Constructing the graph $G'_p$ for ranking URLs $\{u_1, u_2\}$ with respect to the source node $p$. $P_1, P_2, ..., P_5$ are nodes taken from $G_U=(V,E)$ that are reachable from $p$, and $t$ is the super sink. Weights are assigned to edges accordingly, and the maximum flow from $p$ to $t$ is used to rank $u_1$ and $u_2$. }
	\label{fig:flow_construction}
\end{figure}

where $c_e$ is the assigned flow for the edge $e=(v_i,v_j)$ between two nodes, and $c'_e$ is the assigned flow for the edge $e'=(v_i, u_j)$ for the node $v_i$ and the URL $u_j$. The construction of the graph $G'_{p}$ is illustrated in Fig. \ref{fig:flow_construction}. Polynomial running time algorithms such as the Edmonds-Karp algorithm $O(V'E'^{2})$ for finding the maximum flow can be found in \cite{goldberg:1990} and \cite{kleinberg:book}. 

\section{Data Collection}
\label{data-collection}
We collected data from two networks on the web. The first one is the Google Buzz, a platform that combines social relationships and mini-blogging for information dissemination. The second network is Twitter where users choose to follow sources of information. These two networks have messages containing URLs that provide us clues into how users would rank the quality of the information coming from the URLs by using the three techniques we described in Section \ref{algorithms}. 

We collected the Google Buzz data from early September of 2011 to the middle of October of the same year. There were around 2.5M users who shared approximately 100M messages of which about 30M messages had URLs embedded in them. We collected the Twitter data from early September of 2011 to the late December of that year. There were around 1M users who shared approximately 300M messages of which 50M messages had URLs embedded in them. Additional details of the datasets for Google Buzz and Twitter are provided in the Tables \ref{buzz_data_summary} and \ref{twitter_data_summary}. Please note that all URLs refer to all representations of URLs embedded into messages and two different representations could be the same URL when they are masked by redirect services. *URLs refer to the final destination of URLs that have been shared by at least two users within the network.

\subsection{Data Limitations}
First, using Google Buzz and Twitter limits users' demographics which probably is not a representative sample of the entire population as mentioned by authors in \cite{mislove-2011-twitter}. Second, parsing URLs from messages is prone to errors where humans have multiple ways of writing supposedly the same link. Examples are URLs containing typos and spelling mistakes, masked by redirect services, and so on. Third, researchers in \cite{DBLP:kurant} have argued that BFS sampling of a network by starting at a seed generates a large connected component but causes skewness in degree centralities and higher degree averages than in the entire network. 

With limits on hardware resources, bandwidth sharing and data access, we attempted to collect as much as we could for the purpose of ranking URLs on social media. We were able to collect the entire connected component with BFS sampling for Google Buzz, which resulted in the sum of indegree being equal to the sum of outdegree. Twitter is a much larger network that consists of hundreds of millions of accounts. When calculating the data summary of Twitter, we look at users who have been processed in terms of collecting their information and not users who are waiting to be processed, which resulted in the sum of indegree not being equal to the sum of outdegree.

\begin{table}
	\center
	\caption{Data Summary of Google Buzz}
	\label{buzz_data_summary}
	\begin{tabular}{|c|ccc|}
		\hline
		 	&  $\bar{x}$ & $\sigma_X$  & $\sum_{}$  \\
		\hline
		Users  & $-$& $-$& 2,522,109 \\
		Inlinks   & 7.36 & 115.04 & 18,566,607 \\
		Outlinks  & 7.36 & 58.39 & 18,566,607 \\
		Messages   & 42.94 & 1,067.21 & 108,439,019 \\
		All URLs   & 11.67 & 21,706.36 & 34,472,205 \\
		*URLs & 3.85 & 174.80 & 2,647,561 \\
		\hline
	\end{tabular}
		
	\caption{Data Summary of Twitter}
	\label{twitter_data_summary}
	\begin{tabular}{|c|ccc|}
		\hline
		 	&  $\bar{x}$ & $\sigma_X$  & $\sum_{}$  \\
		\hline
		Users  & $-$ & $-$ & 1,057,163 \\
		Inlinks   & 17,675.58 & 334,127.10 & 18.69B \\
		Outlinks  & 520.66 & 7,676.48 & 550,421,023 \\
		Messages   & 280.84 & 1,005.09 & 277,310,683 \\
		All URLs   & 44.26 & 45,359.19 & 46,532,403 \\
		*URLs 	   & 8.19 & 57.59 & 2,294,077\\
		\hline
	\end{tabular}
\end{table}

\subsection{Data Analysis}
Two sets of URLs are considered for the purpose of our data analysis. From both Google Buzz and Twitter datasets, we have randomly chosen 2,000 URLs with equal probability denoted as the random set of URLs. We also have chosen the top 2,000 shared URLs denoted as the popular set of URLs. There are two sets of URLs in each network giving us four sets of URLs in total. For each URL, we calculated the size of the affected set consists of nodes that received the URL from the spreaders but chose not to spread it further. 

\begin{figure}[htp!]
  \center
  \subfigure[Google Buzz]{\includegraphics[scale=.29]{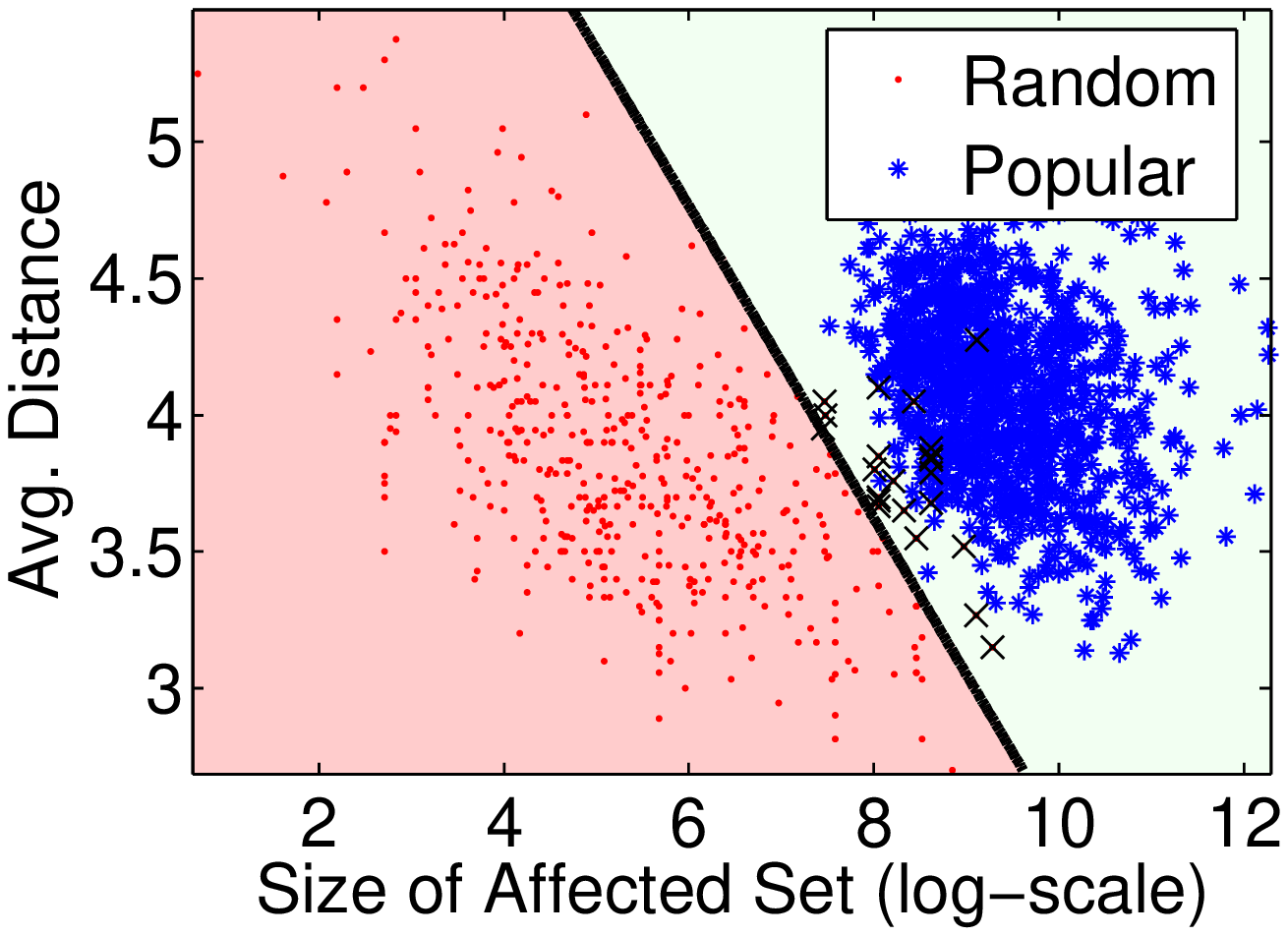}}
  \subfigure[Twitter]{\includegraphics[scale=.29]{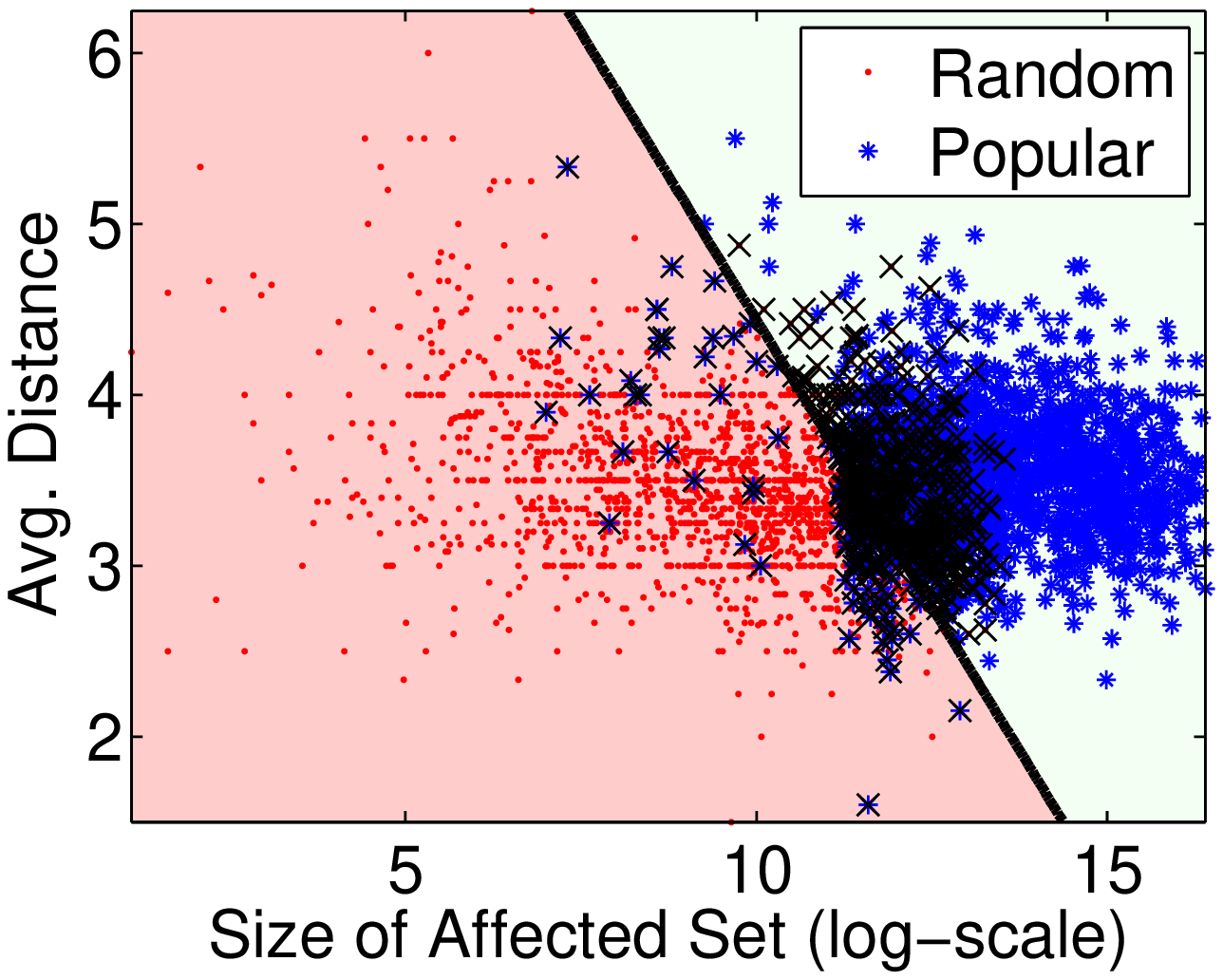}}
  \caption{A point on the plot is a URL where the x-axis corresponds to the size of the affected set in logarithmic scale, and the y-axis corresponds to the average length of shortest paths from randomly chosen users to the spreaders. A red point is a URL from the random set, and a blue star is a URL from the popular set. The black line is a linear classifier that separates popular URLs from random URLs and crosses are points that have been miss-classified.}
  \label{fig:info-distances}
\end{figure}

We also computed the average length of all shortest paths from 10 randomly chosen users to members of a random subset of spreaders. The results are shown in Fig. \ref{fig:info-distances}(a) for Google Buzz and Fig. \ref{fig:info-distances}(b) for Twitter. We substitute the entire spreader set with a randomly selected subset simply as a matter of efficiency because shortest-path computations are expensive in large networks as mentioned by authors in \cite{DasSarma}.

In Fig. \ref{fig:info-distances}, we noticed that as the size of the affected set increases, the average distance from randomly selected users to the information on the web page decreases for random and popular sets of URLs in Google Buzz. This is because very large affected sets increase the likelihood that a randomly chosen user has a path through an affected user reaching a spreader. This agrees with our intuition that information collectively shared by users with high outdegrees has a greater coverage of dissemination.  However, this correlation is weaker in Twitter due to the celebrity effect of some users having millions of followers and creating large affected sets. For instance, a URL that was only shared in the network by a celebrity. More importantly, affected sets influence our social ranking techniques where the structure of the network instead of the web topology is used to rank pages or URLs. For example, the PageRank on a social network (PRSN) would rank URLs that were shared by high outdegree spreaders higher because they absorb most of the scores distributed to them. Our maximum flow approach to personalize social ranking would be affected at the first level if a user directly follows a high outdegree spreader. Because of the celebrity effect in Twitter, this rank increase will also carry over the subsequent levels because the scores could be circulated to the rest of the network by the intricate social relationships. Interestingly, HITS is not affected by the network structure since the algorithm does not consider social relationships but only takes into account which person shares what URL.
 
\section{Social Ranking Experiments}
\label{search_experiment}
For each network, we selected 30 URLs from the popular and random URLs sets. For each selected URL, we calculated its score by using PageRank and HITS, and ranked the URLs (i.e, 1st, 2nd, 3rd, etc.) with respect to the set. We also ranked the selected URLs tailored to four randomly chosen users using maximum flow. Results are shown in Table \ref{table:buzz_pop_urls} for popular URLs in Google-Buzz where we enumerated the 30 selected URLs in the first column, ranking positions using PageRank in the second column, HITS in the third column, and maximum flow in the fourth column. In the fourth column, the first element corresponds to the first person, second element corresponds to the second person, and so on. We did the same for the random set of URLs in Google Buzz shown in Table \ref{table:buzz_rand_urls}. The ranking results of Twitter are not shown as a full table, and full representations of the URLs listed in these tables have been shorten to save space.

\begin{figure*}[htp!]
  \center
  \subfigure[Twitter Popular URLs.]{\includegraphics[scale=.50]{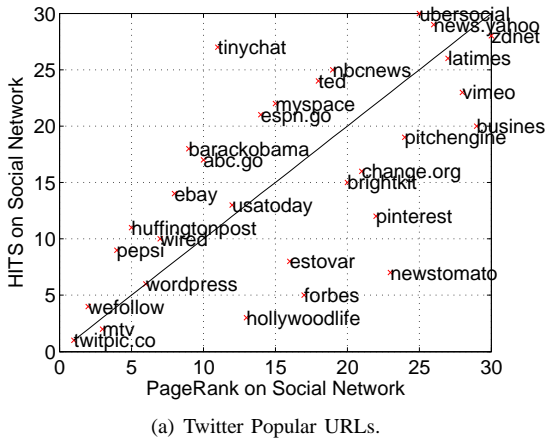}}
  \subfigure[Twitter Random URLs. ]{\includegraphics[scale=0.50]{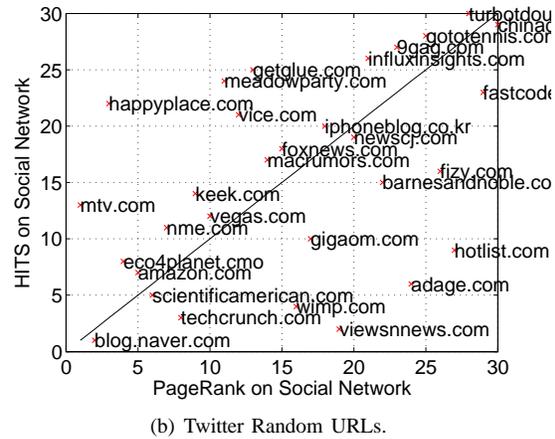}}
  \caption{A point is a URL where the x-axis corresponds to the ranking position determined by PageRank and the y-axis corresponds to the ranking position determined by HITS. A value of 1, 2, 3, etc. denotes the first, second, third, etc. position of the URL. When a URL lies on the $y=x$ line, then both the PageRank and HITS algorithm ranked the URL in the same position.}
  \label{fig:pagerank_vs_hits}
\end{figure*}

We compared the ranking results of PageRank and HITS shown in Fig. \ref{fig:pagerank_vs_hits} for Twitter. Ranking Results of Google Buzz are listed in Table \ref{table:buzz_pop_urls} and \ref{table:buzz_rand_urls}. The ranking of popular URLs using PageRank and HITS are more consistent than the random URLs. We measured the ranking consistency as the average difference of two ranking algorithms on a set of URLs (i.e., $\frac{1}{w}\sum_{u \in U'} |P_{HSN}(u) - P_{PRSN}(u)|$) and the sum of differences (i.e., $\sum_{u \in U'} |P_{HSN}(u) - P_{PRSN}(u)|$) where $P_{x}(u)$ is the position of the URL $u$ determined by the algorithm $x$ and $w$ is the number of URLs. 

For the popular URLs in Google Buzz, the average difference was 2.9 meaning that on average HITS and PageRank were off by 3 positions and the sum of differences between them was 86. For the random URLs in Google Buzz, the average difference was 9.6 and the sum of differences between them was 288. For the popular URLs in Twitter, the average difference was 5.9 and the sum of differences between them was 178. For random URLs in Twitter, the average difference was 7.2 and the sum of differences between them was 216. In both networks, popular URLs are ranked more consistently than random URLs which makes the HITS algorithm more suitable than PageRank when ranking viral information because it is computationally more efficient. 

\begin{figure*}[htp!]
  \center
  \subfigure[Popular URLs.]{\includegraphics[scale=0.30]{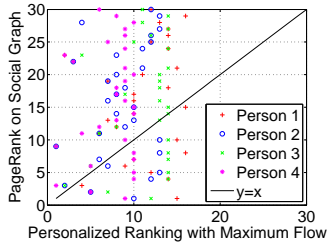}}
  \subfigure[Popular URLs. ]{\includegraphics[scale=0.30]{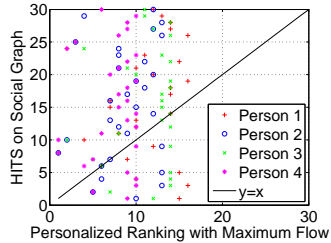}}
  \subfigure[Random URLs. ]{\includegraphics[scale=0.30]{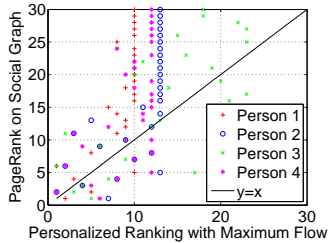}}
  \subfigure[Random URLs. ]{\includegraphics[scale=0.30]{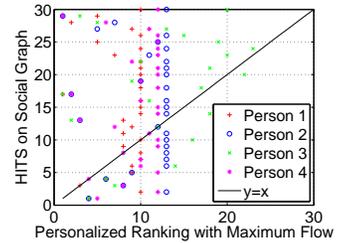}}
  \caption{Social Ranking with Four Randomly Selected Users on Google Buzz. }
  \label{fig:buzz_social_rank}
\end{figure*}

\begin{figure*}[htp!]
  \center
  \subfigure[Popular URLs.]{\includegraphics[scale=0.30]{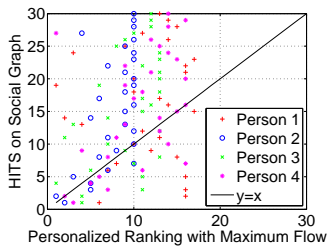}}
  \subfigure[Popular URLs. ]{\includegraphics[scale=0.30]{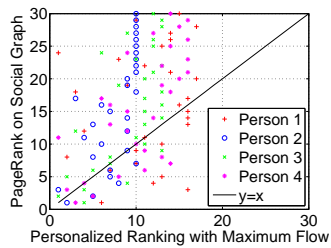}}
  \subfigure[Random URLs. ]{\includegraphics[scale=0.30]{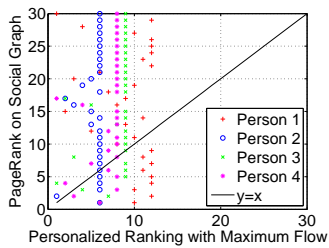}}
  \subfigure[Random URLs. ]{\includegraphics[scale=0.30]{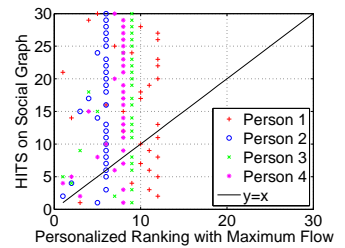}}
  \caption{Social Ranking with Four Randomly Selected Users on Twitter.}
  \label{fig:twitter_social_rank}
\end{figure*}

We noticed that the ranking results determined by each individual user using maximum flow are less correlated with themselves than the results computed by PageRank and HITS. First, we compared the ranking results of maximum flow with PageRank and HITS using popular and random URLs for Google Buzz shown in Fig. \ref{fig:buzz_social_rank}. The first and second plots on the left are ranking results of popular URLs and the third and fourth plots on the right are ranking results of random URLs labelled by their sub-captions. A point on the graph is a URL where the x-axis is the ranking position of the URL determined by maximum flow and the y-axis is the ranking position determined by either PageRank or HITS labelled on the y-axis. The identical layout for Twitter is shown in Fig. \ref{fig:twitter_social_rank}.

\begin{table}[htbp]
  \centering
  \caption{Ranking Results of 30 Popular URLs in Google Buzz}
    \begin{tabular}{|r|r|r|r|}
    \hline
    \multicolumn{1}{|c|}{URLs } & \multicolumn{1}{c|}{PRSN} & \multicolumn{1}{c|}{HSN} & \multicolumn{1}{c|}{MF} \\
    \hline
    \multicolumn{1}{|l|}{abcnews.go} & 1     & 1     & 9/12/10/15 \\
    \multicolumn{1}{|l|}{youtube} & 2     & 2     & 5/7/5/6 \\
    \multicolumn{1}{|l|}{yahoo} & 3     & 10    & 1/2/2/4 \\
    \multicolumn{1}{|l|}{businessweek} & 4     & 14    & 10/14/12/14 \\
    \multicolumn{1}{|l|}{bloomberg} & 5     & 9     & 10/14/13/12 \\
    \multicolumn{1}{|l|}{wordpress} & 6     & 7     & 5/5/7/9 \\
    \multicolumn{1}{|l|}{nytimes} & 7     & 4     & 10/14/6/10 \\
    \multicolumn{1}{|l|}{appleinsider} & 8     & 3     & 10/14/13/16 \\
    \multicolumn{1}{|l|}{facebook} & 9     & 8     & 1/1/1/1 \\
    \multicolumn{1}{|l|}{wired} & 10    & 5     & 9/14/13/15 \\
    \multicolumn{1}{|l|}{lockerz} & 11    & 6     & 4/6/6/6 \\
    \multicolumn{1}{|l|}{apple} & 12    & 11    & 6/8/9/8 \\
    \multicolumn{1}{|l|}{pcworld} & 13    & 15    & 8/13/10/7 \\
    \multicolumn{1}{|l|}{guardian} & 14    & 12    & 10/14/8/10 \\
    \multicolumn{1}{|l|}{reuters} & 15    & 19    & 10/14/10/16 \\
    \multicolumn{1}{|l|}{ted} & 16    & 13    & 9/13/7/10 \\
    \multicolumn{1}{|l|}{amazon} & 17    & 21    & 8/9/8/10 \\
    \multicolumn{1}{|l|}{techcrunch} & 18    & 17    & 8/13/9/14 \\
    \multicolumn{1}{|l|}{engadget} & 19    & 16    & 9/13/7/7 \\
    \multicolumn{1}{|l|}{reddit} & 20    & 23    & 10/13/8/11 \\
    \multicolumn{1}{|l|}{empireavenue} & 21    & 22    & 9/14/11/15 \\
    \multicolumn{1}{|l|}{boston} & 22    & 25    & 3/3/3/3/ \\
    \multicolumn{1}{|l|}{xkcd} & 23    & 24    & 2/4/8/2 \\
    \multicolumn{1}{|l|}{whitehouse} & 24    & 18    & 9/14/11/14 \\
    \multicolumn{1}{|l|}{gizmodo} & 25    & 20    & 7/10/12/12 \\
    \multicolumn{1}{|l|}{pingchat} & 26    & 27    & 9/12/12/14 \\
    \multicolumn{1}{|l|}{thesocialnetwork-movie} & 27    & 28    & 9/14/13/14 \\
    \multicolumn{1}{|l|}{bbc} & 28    & 29    & 10/11/4/13 \\
    \multicolumn{1}{|l|}{photofocus} & 29    & 26    & 8/14/13/16 \\
    \multicolumn{1}{|l|}{stackoverflow} & 30    & 30    & 6/11/12/12 \\
	\hline
    \end{tabular}%
  \label{table:buzz_pop_urls}%
\end{table}%

\begin{table}[htbp]
  \centering
  \caption{Ranking Results of 30 Random URLs in Google Buzz}
    \begin{tabular}{|r|r|r|r|}
	\hline
    \multicolumn{1}{|c|}{URLs } & \multicolumn{1}{c|}{PRSN} & \multicolumn{1}{c|}{HSN} & \multicolumn{1}{c|}{MF} \\
    \hline
    \multicolumn{1}{|l|}{networkedblogs} & 1     & 28    & 6/5/7/2 \\
    \multicolumn{1}{|l|}{picasaweb.google} & 2     & 29    & 1/3/1/5 \\
    \multicolumn{1}{|l|}{ping.fm} & 3     & 1     & 5/4/4/4 \\
    \multicolumn{1}{|l|}{thenextweb} & 4     & 3     & 8/7/8/3 \\
    \multicolumn{1}{|l|}{twitter} & 5     & 18    & 12/17/13/10 \\
    \multicolumn{1}{|l|}{income4free} & 6     & 17    & 2/1/2/1 \\
    \multicolumn{1}{|l|}{fastestwaylosebellyfat} & 7     & 19    & 10/9/10/10 \\
    \multicolumn{1}{|l|}{digg} & 8     & 25    & 12/19/12/5 \\
    \multicolumn{1}{|l|}{sports.espn.go} & 9     & 4     & 4/6/6/6 \\
    \multicolumn{1}{|l|}{wired} & 10    & 5     & 12/21/9/9 \\
    \multicolumn{1}{|l|}{businessinsider} & 11    & 13    & 3/2/3/8 \\
    \multicolumn{1}{|l|}{forbes} & 12    & 12    & 7/12/12/9 \\
    \multicolumn{1}{|l|}{foxnews} & 13    & 27    & 11/13/5/9 \\
    \multicolumn{1}{|l|}{behance} & 14    & 11    & 11/23/13/8 \\
    \multicolumn{1}{|l|}{huffingtonpost} & 15    & 23    & 12/20/11/7 \\
    \multicolumn{1}{|l|}{entrepreneur} & 16    & 2     & 12/21/13/10 \\
    \multicolumn{1}{|l|}{puntogov} & 17    & 15    & 12/23/13/10 \\
    \multicolumn{1}{|l|}{addictivefonts} & 18    & 6     & 10/14/13/9 \\
    \multicolumn{1}{|l|}{theprism} & 19    & 30    & 12/20/13/10 \\
    \multicolumn{1}{|l|}{telegraph} & 20    & 22    & 9/10/13/10 \\
    \multicolumn{1}{|l|}{npr} & 21    & 7     & 10/19/13/10 \\
    \multicolumn{1}{|l|}{popsci} & 22    & 16    & 10/11/13/10 \\
    \multicolumn{1}{|l|}{economist} & 23    & 10    & 12/16/13/10 \\
    \multicolumn{1}{|l|}{marketwatch} & 24    & 8     & 8/8/13/10 \\
    \multicolumn{1}{|l|}{opencog} & 25    & 9     & 12/23/13/8 \\
    \multicolumn{1}{|l|}{dslreports} & 26    & 26    & 12/15/13/10 \\
    \multicolumn{1}{|l|}{last.fm} & 27    & 24    & 12/23/13/10 \\
    \multicolumn{1}{|l|}{tech.slashdot} & 28    & 20    & 12/22/13/10 \\
    \multicolumn{1}{|l|}{wimp} & 29    & 21    & 12/18/13/10 \\
    \multicolumn{1}{|l|}{socialturns} & 30    & 14    & 12/18/13/10 \\
  	\hline
    \end{tabular}%
  \label{table:buzz_rand_urls}%
\end{table}%

For personalized ranking, we measured the ranking consistency as the average difference of a pair of users with respect to a URL set. For instance, in the Table \ref{table:buzz_social_ranking}, the left column and the top row are the four selected users where the element $a_{ij}$ corresponds to the average difference of users $i$ and $j$. Please note the upper triangle or elements above the diagonal refer to the random URLs and the lower triangle or elements below the diagonal refer to the popular URLs. The right column refers to the outdegree of users in the random URLs, and the last row refers to the outdegree of users in the popular URLs. For Twitter, the ranking results in the same format are given in Table \ref{twitter_social_ranking}.  

For random URLs in Google Buzz, we noticed that persons $p_1$ and $p_3$ have an average difference of 1.7 where $p_2$ and $p_4$ have an average difference of 6.7. For popular URLs, the variability is smaller where $p_4$ and $p_2$ have an average difference of 2.0 and $p_1$ and $p_2$ have an average difference of 3.2. Outdegree measures the number of people a user follows since the ranking results are based on them. And finally, ties are expected when using maximum flow since the number of URLs shared among friends is minuscule compare to the number of pages in the deep Web. Therefore, we simply use PageRank or HITS to break ties among pages when necessary. 

\begin{table}[htbp!]
  \centering
  \caption{Avg. Ranking Differences in Google Buzz}
    \begin{tabular}{|r|rrrr|r|}
    \hline
    -     & $p_1$    & $p_2$    & $p_3$    & $p_4$ & outdegree.\\
    \hline
    $p_1$    & -     & 5.1   & 1.7   & 2.4 	&	369\\
    $p_2$    & 3.2   & -     & 4.8   & 6.7 	&	4,505\\
    $p_3$    & 2.5   & 2.6   & -     & 3.1  &	1,125\\
    $p_4$    & 3.2   & 2.0   & 2.5   & -    &	102\\
    \hline
    out deg.	 & 159	 & 355	 &503	&340 &\\
    \hline
    \end{tabular}
  \label{table:buzz_social_ranking}%
\end{table}%

\begin{table}[htbp!]
  \centering
  \caption{Avg. Ranking Differences in Twitter}
    \begin{tabular}{|r|rrrr|r|}
    \hline
     -     & $p_1$    & $p_2$    & $p_3$    & $p_4$ & outdegree.\\
    \hline
    $p_1$    & -      & 1.5   & 2.0   & 4.0 	&	203\\
    $p_2$    & 3.7    & -     & 3.0   & 3.8 	&	122\\
    $p_3$    & 3.3    & 3.3  & -     & 4.6 		&	426\\
    $p_4$    & 3.7   & 3.8  & 5.2   & -  		&	119\\
    \hline
    out deg. & 324	  & 158  &129	  &1,731 &\\
    \hline
    \end{tabular}%
  \label{twitter_social_ranking}%
\end{table}%

\section{Related Work}
\label{related_work}
Our work lies at the intersection of the study of social network analysis and the ranking techniques in information retrieval. The closest to our work are references \cite{bahmani} \cite{bao} \cite{carmel} in which the authors studied the problem of social searching while we studied the problem of social ranking. In \cite{bahmani}, authors proposed an approximation to an algorithm called Partitioned Multi-Indexing to rank queries on the content generated in social networks by using a distributed hash table and schemas for updating the content continuously generated by the users. One similarity is that both theirs approach and ours consider information shared by social ties to be an important element in searching and ranking. Still, their work approximates network distances between users while our work uses the maximum flow of a constructed network. Another difference is that we do not focus on answering queries with social ties but on designing ranking techniques of URLs which could be used to answer friendship-related queries. In \cite{bao}, authors proposed simple techniques to re-rank search results based on Similarity and Familiarity networks using their enterprise social network. 

While social searches have been introduced in multiple settings from the Social Query Model (SQM) \cite{banerjee} to the implementation of social searching applications for mobile devices \cite{aardvark}, a good amount of work has focused on finding the right answer to a search query by routing the search query to the right person in a social network graph \cite{ilink}\cite{aardvark}.  We studied the structure of the network to socially and automatically rank URLs without users intervention. In the Social Query Model \cite{banerjee}, routing paths of search queries are studied in decentralized systems where indeterministic behavior of each agent willing to provide a correct answer with some level of accuracy and expertise is taken into consideration when forming an optimal routing policy. In Aardvark \cite{aardvark}, the focus was to route a query from the searcher to a designated user in a social network that was assumed to be able to provide an answer. We took the approach of using network flow where the goal is to automatically rank a set of pages through the eyes of the searcher's social ties.

Indegree-based algorithms such as PageRank \cite{Brin:1998}, SALSA \cite{salsa}, and HITS \cite{Kleinberg-hits} are used for ranking pages on a web graph where an edge between two pages represents an endorsement of one page by another page. The intuition behind network flow is that it automatically incorporates indegree analysis where a node that does not share a web page will distribute its flow to the sources that it follows, and sources of high indegree will eventually get the largest share of flow if the information is not found locally. In \cite{bao}, authors looked at direct annotations from users in Delicious to enhance searches while we look at shared messages embedded with URLs to rank pages. To the best of our knowledge, we are the first to propose using maximum flow to personalize the ranking of pages based on the messages containing URLs that users share in online social networks. 

\section{Conclusion}
\label{conclusion}
Information shared between users in online social networks such as URLs provides a unique perspective of the ranking of pages. In our approach, humans instead of pages are the ones who rank the URLs by sharing them, and the social network of the users instead the web graph topology is used to propagate the ranking. 

First, we collected two large-scale information networks of online users to study how users in these networks share URLs which impacts the distance between a person and a URL. For instance, researchers in \cite{albert:1999} estimated the number of hops between any two pages to be on average 19; while Milgram estimated that the number of hops between any two people is no more than 6 \cite{milgram67}. Since information propagates differently in social networks, the social structure bounds how far a person is away from a shared URL. 

Second, we reinterpreted the ranking techniques of PageRank and HITS and proposed to use maximum network flow to personalized the ranking of pages tailored to each individual user. Maximum flow detects the popularity of a shared URL among friends but popularity does not necessary reflect endorsement. We expected that each unique individual would rank the URLs differently, since no two people on a social network are the same. Interestingly, the ranking results of popular URLs using PageRank and HITS are more correlated than random URLs suggesting that the overall view of users on ubiquitous information is more consistent, but everyone has their own opinion in the end. Instead of attempting to socially rank the entire web, we re-ranked a selected set of URLs to make it scalable and efficiently executable for search engines. If the size of the web doubles in the next few years, it would not affect our approach since only a subset of URLs that users shared are actually re-ranked. 

More importantly, we believe that personalizing the ranking is useful for social searching because it provides a mechanism for the interaction between the searcher and the sharer where the searcher can discuss with the sharer about the item relating to a query on a search engine. For instance, a new product that the sharer posted on appleinsider.com or a piece of political news on nytimes.com. This potential interaction between the searcher and the sharer is valuable because the influence of the sharer on the searcher is stronger than the influence coming from the authorities detected by HITS and PageRank in many non-technical and social situations but not for all. This feature could be implemented in search engines where pages returned to a given query are re-ranked via social networks if there are pages shared among friends or other associates of the searcher that are related to the query. 

\bibliography{mybib}

\section{Acknowledgement} 
Research was sponsored by the Army Research Laboratory and was accomplished under Cooperative Agreement Number W911NF-09-2-0053. The views and conclusions contained in this document are those of the authors and should not be interpreted as representing the official policies, either expressed or implied, of the Army Research Laboratory or the U.S. Government. The U.S. Government is authorized to reproduce and distribute reprints for Government purposes notwithstanding any copyright notation here on.
\end{document}